# Unravelling Metamaterial Properties in Zigzag-base Folded Sheets

Maryam Eidini,[1]* Glaucio H. Paulino[1]

[1]Department of Civil and Environmental Engineering, University of Illinois at Urbana Champaign, 205 North Mathews Ave., Urbana, IL 61801, USA.

*Correspondence to: eidinin1@illinois.edu

## Abstract

Creating complex spatial objects from a flat sheet of material using origami folding techniques has attracted attention in science and engineering. In the present work, we employ geometric properties of partially folded zigzag strips to better describe the kinematics of the known zigzag/herringbone-base folded sheet metamaterials such as the Miura-ori. Inspired by the kinematics of a one-degree of freedom zigzag strip, we introduce a class of cellular folded sheet mechanical metamaterials comprising different scales of zigzag strips in which the class of the patterns combines origami folding techniques with kirigami. Employing both analytical and numerical models, we study the key mechanical properties of the folded materials. Particularly, we show that, depending on the geometry, these materials exhibit both negative and positive in-plane Poisson's ratio. By expanding the design space of the Miura-ori, our class of patterns is potentially appropriate for a wide range of applications, from mechanical metamaterials to deployable structures at both small and large scales.

## Introduction

Origami, the art of paper folding, has been a substantial source of inspiration for innovative design of mechanical metamaterials [1, 2, 3, 4], for which the material properties arise from their geometry and structural layout. Kirigami is the art of paper cutting and, for engineering applications, it has been used as combination of origami with cutting patterns to fabricate complex microstructures through micro-assembly [5], as well as to create 3D core structures [6, 7, 8, 9]. Furthermore, rigid origami is a subset of origami structures where rigid panels (facets) are linked through perfect hinges leading to an entirely geometric mechanism. The mathematical theory of rigid origami has been studied by various researchers [10, 11, 12, 13, 14]. Based on rigid origami behavior of folding patterns, recent research [1, 3] has shown that in Miura-ori, metamaterial properties arise due to their folding geometry. Miura-ori is a classic origami folding pattern and its main constituents are parallelogram facets, which are connected along fold lines. Morphology and/or mechanisms similar to that of Miura-ori naturally arises in insect wings [15], tree leaves [16] and embryonic intestine [17, 18]. Moreover, it has been reported that self-organized wrinkling pattern for a planar stiff thin elastic film connected to a soft substrate subjected to biaxial compression is a herringbone pattern which is similar to the Miura-ori pattern [19, 20, 21]; and that the herringbone pattern corresponds to the minimum energy configuration [22]. Applications of the Miura-ori pattern range from folding of maps [23] to technologies such as deployable solar panels [24], foldcore sandwich panels [25, 26], and metamaterials [1, 3].

Motivated by outstanding properties and broad range of applications of the Miura-ori, the present study starts with raising a question: Can we design patterns aiming at both preserving the remarkable properties of the Miura-ori and expanding upon its design space? In this regard, upon closer inspection of the Miura-ori, we associate its kinematics to that of a one-degree of freedom (DOF) zigzag strip, and we present a technique to create zigzag-base mechanical metamaterials including various scales of zigzag strips. Through this study, we answer the question positively.

An important material property used to create the patterns and to study the size change of the folded sheets in the present work, is the Poisson's ratio. It is defined as the negative ratio of elastic extensional strain in the direction normal to the applied load, to the axial extensional strain in the direction of the applied load. Most commonly, when a material is stretched in a given direction it tends to get narrower in the directions perpendicular to the applied load. However, when stretched, materials with negative Poisson's ratio or auxetic materials expand in the direction perpendicular to the applied load. Under bending, anticlastic (saddle-shape) curvature and synclastic (spherical-shape) are observed in materials with positive and negative Poisson's ratios, respectively [27, 28]. Based on the theory of elasticity, the Poisson's ratio for a thermodynamically stable isotropic linear elastic material is bounded between -1 to 0.5 [29]. Contrary to isotropic solids, the value of Poisson's ratio is unrestricted in an anisotropic elastic material ($-\infty < \upsilon < \infty$) [30]. Folded sheets are anisotropic in which the deformation happens due to folding and unfolding. Thus, in folded sheet materials (for instance in most ranges of geometric parameters of the Miura-ori folding pattern), the Poisson's ratio range can be beyond the bound for isotropic materials [1, 3].

## Kinematics of a folded one-DOF zigzag strip

In this section, we look closely at kinematics of the Miura-ori as a zigzag-base folding pattern, which provides inspiration to create a class of mechanical metamaterials. A regular Miura-ori sheet contains zigzag strips of parallelogram facets, in which each unit cell can be decomposed into two *V*-shapes (Fig. 1A). Each *V*-shape includes 2 rigid parallelogram facets connected via a hinge along the joining ridges as shown in Fig. 1B. We show that the kinematics of a properly constrained *V*-shape, as described below, is a function of an angle in the horizontal *xy*-plane. The constraints on the *V*-shape are applied to simulate similar conditions to those of the *V*-shapes in the Miura-ori sheet, i.e., to create a one-DOF planar mechanism. Hence, to remove the rigid body motions associated with the translational and rotational DOFs, we constrain all translational DOFs of the point *A* and assume that the edges *AB* and *BC* of the facets move in the *xy*-plane, and that the projected length of the edge *AD* in the *xy*-plane remains along the *x*-axis (i.e., point *D* moves in the *xz*-plane). With this set up, the *V*-shape has only one planar DOF. The expressions defining the geometry of the *V*-shape are given by

$$\ell_v = a \frac{\cos \alpha}{\cos \phi} \qquad w_v = 2b \sin \phi \qquad (1)$$

where $\ell_v$ is the projected length of the edges $a$ in the **xy**-plane and in the **x** direction; $w_v$ is the width of the semi-folded *V*-shape in the **xy**-plane and along the **y** direction; $\phi$ is an angle in the **xy**-plane between the edge $b$ and the **x**-axis. The in-plane Poisson's ratio of the system is given by

$$(\upsilon_{w\ell})_V = -\frac{\varepsilon_{\ell_v}}{\varepsilon_{w_v}} = -\frac{d\ell_v/\ell_v}{dw_v/w_v} = -\tan^2\phi \quad (2)$$

The above expression shows that the kinematics of a *V*-shape is only a function of the angle $\phi$. In particular, it shows that in a unit cell containing two *V*-shapes arranged side-by-side in the crease pattern, we can scale down the length $b$ of the parallelogram facets to $1/n$ that of the other joining *V*-shape ($n$ is a positive integer), while preserving the capability of folding and unfolding. Using this insight in our current research, we created a class of zigzag-base metamaterials in which the unit cell includes two different scales of *V*-shapes possessing the same $\phi$ angles (Fig. 1C). For instance, for the zigzag strips shown in Fig. 1C, the value of $n$ is equal to 2. For the case of $n=2$, from both the numerical model and the geometry, the ideal unit cell has only one planar mechanism (see Section (6-1) in Appendix), i.e., the geometry of the unit cell constrains the *V*-shapes properly to ideally yield one-DOF planar mechanism. Therefore, the condition for which we have obtained the kinematics of the *V*-shape is met.

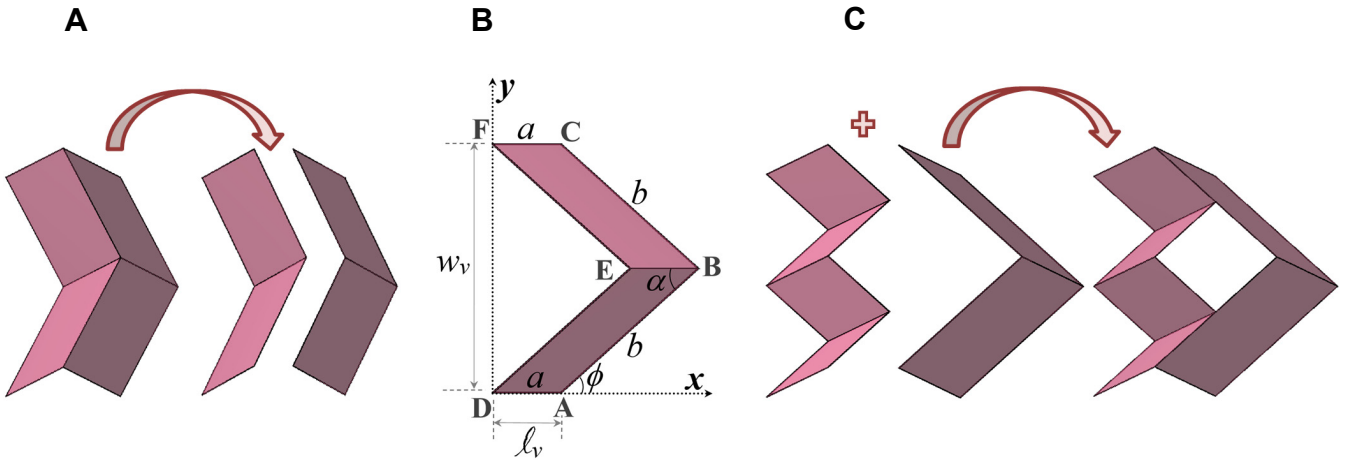

**Fig. 1. From Miura-ori to zigzag-base foldable metamaterials possessing different scales of zigzag strips.** (**A**) A Miura-ori unit cell contains two *V*-shapes aligned side-by-side forming one concave valley and three convex mountain folds (or vice versa, if the unit cell is viewed from the opposite side). (**B**) Top view of a *V*-shape fold including two identical parallelogram facets connected along the ridges with length $a$. Its geometry can be defined by the facet parameters $a$, $b$, $\alpha$ and the angle $\phi \in [0, \alpha]$. (**C**) Two different scales of *V*-shapes with the same angle $\phi$ are connected along joining fold lines. The length $b$ of the parallelogram facets in the left zigzag strip of *V*-shapes is half that of the strip on the right in the unit cell shown.

## BCH$_n$ Zigzag-base patterns

The *Basic unit Cell with Hole* (BCH$_n$) of the patterns introduced in Fig. 1C is parametrized in Fig. 2A. The unit cell includes 2 large and $2n$ small parallelogram rigid panels joined via fold lines. For example, for the unit cell shown in Fig. 2A, $n$ is equal to 2. Although theoretically possible, large values of $n$ have not been explored. For a large $n$, a zigzag strip of small parallelograms approaches a narrow strip. In current research, we use only $n=2, 3$ in the BCH patterns with an emphasis placed on $BCH_2$. We can define the unit cell by the geometry of the parallelogram facets with sides $a$ nd $b$ and acute angle $\alpha \in [0, \pi/2]$, and angle $\phi \in [0, \alpha]$, which is half the angle between the edges $b_1$ in the **xy**-plane. The expressions defining the geometry of the $BCH_n$ are given by

$$w = 2b\sin\phi \qquad \ell = 2a\frac{\cos\alpha}{\cos\phi} \qquad h = a\sin\alpha\sin\theta \qquad b_1 = b/n \qquad (3)$$

Note that $\ell$ is the projected length of the zigzag strips along the **x**-axis in the **xy**-plane. The relationship between the angle $\phi$ and the fold angle $\theta$ is as follows:

$$\tan\phi = \cos\theta\tan\alpha \qquad (4)$$

The outer dimensions of a sheet made of tessellation of the same $BCH_n$ (Fig. 2B) are given by

$$W = m_2(2b\sin\phi) \qquad L = m_1\left(2a\frac{\cos\alpha}{\cos\phi} + \frac{n-1}{n}b\cos\phi\right) + \frac{1}{n}b\cos\phi \qquad (5)$$

Note that, in the relation given for the length $L$, the expression within the parentheses presents the length of the repeating unit cell, and the last term shows the edge effect. The second term within the parentheses is related to the effect of the holes in the tessellation.

For the case of rigid panels connected via hinges at fold-lines, from the geometry, the BCH with $n=2$ has only one independent DOF. In general, the number of DOFs for each unit cell of $BCH_n$ is $(2n-3)$. Using at least two consecutive rows of small parallelograms, instead of one, in $BCH_n$ (see Fig. A1B) decreases the DOFs of the BCH to one irrespective of the number of $n$ (see Fig. A2 and Section (6-1) in Appendix for more details). Additionally, the patterns are all rigid and flat-foldable, and can be folded from a flat sheet of material, i.e., they are developable. In Fig. 3, a few configurations of the patterns are presented.

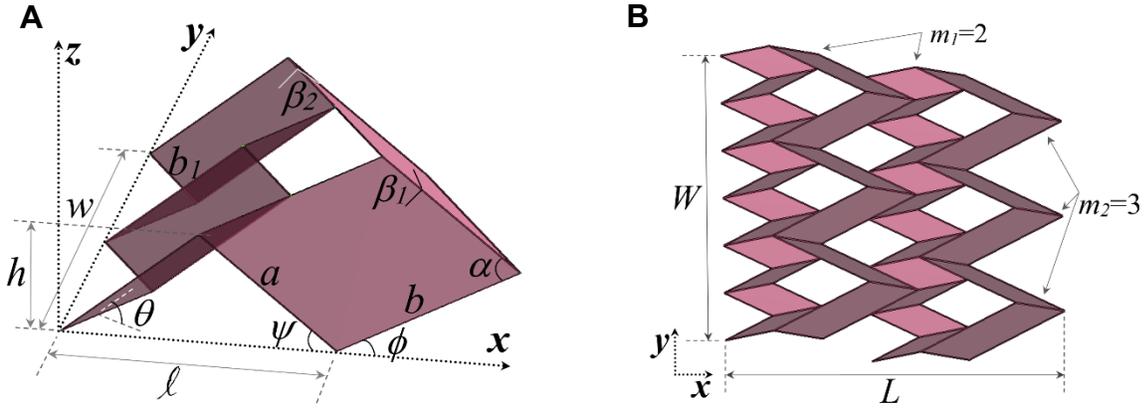

**Fig. 2. Geometry of $BCH_n$ pattern**. (**A**) Geometry of the unit cell. The geometry of a sheet of $BCH_n$ can be parameterized by the geometry of a parallelogram facet ($a$, $b$ and $\alpha$), half number of small parallelogram facets ($n$) and fold angle $\phi \in [0, \alpha]$ which is the angle between fold lines $b$ and the **x**-axis. Other important angles in the figure are fold angle between the facets and the **xy**-plane, i.e., $\theta \in [0, \pi/2]$; angle between the fold lines $a$ and the **x**-axis, i.e., $\psi \in [0, \alpha]$; and dihedral fold angles between parallelograms, $\beta_1 \in [0, \pi]$ and $\beta_2 \in [0, \pi]$, joined along fold lines $a$ and $b$, respectively. (**B**) A sheet of $BCH_2$ with $m_1=2$, $m_2=3$ and outer dimensions of $L$ and $W$.

## Mechanical properties of BCH$_n$ patterns

Two different values have been reported in the literature [1, 3, 31] for the in-plane Poisson's ratio of a partially folded Miura-ori sheet. Associating the Poisson's ratio given in [1, 3] to the kinematics of the folded zigzag-base sheet with one-DOF planar mechanism, and the one suggested in reference [31] to the Poisson's ratio obtained by considering the end-to-end dimension of the sheet, clarifies the issue. In this regard, note that while the first approach is valuable to provide insight on the kinematics of a zigzag-base folded sheet such as Miura-ori and BCH$_2$, the latter definition can also be relevant depending on the application. To place emphasis on these two important concepts in relation to the folded sheet materials introduced in this work, we named the value obtained by the first approach as $\upsilon_z$ and the latter as $\upsilon_{e-e}$, where the indices $z$ and $e$-$e$ stand for zigzag and end-to-end, respectively (see Fig. 4A and Fig. A3). In this regard, for the sheet of BCH$_2$ shown in Fig. 2B, $\ell$ and $L$ are used to obtain $\upsilon_z$ and $\upsilon_{e-e}$, respectively. Notice that $\ell$ for a sheet is the sum of the projected lengths of the zigzag strips in the *xy*-plane and parallel to the *x*-axis, and for a sheet made of tessellation of identical BCH$_2$ is equal to $m_1$ times that of a unit cell (Fig. 4A). Also, $L$ is the end-to-end dimension of the sheet as shown in Fig. 2B. Note that since the width of the sheet along the *y*-axis is always a factor of $\sin\phi$, in both approaches we can consider $W=w$. Hence, $\upsilon_z$ of the sheet is given in the following which is equal to the kinematics of a *V*-shape described in the previous sections.

$$(\upsilon_{w\ell})_z = -\frac{\varepsilon_\ell}{\varepsilon_w} = -\frac{d\ell/\ell}{dw/w} = -\tan^2\phi \tag{6}$$

Accordingly, BCH$_2$ and all other combined patterns of BCH with one-DOF planar mechanism (e.g., patterns shown in Fig. 3) have $\upsilon_z$ equal to $-\tan^2\phi$ (i.e., $\upsilon_z$ of a Miura-ori sheet with the same fold angle (*1, 3*) - see Fig. 4 A and B). We emphasize that the in-plane Poisson's ratio, obtained by considering the projected lengths of the zigzag strips in the patterns, also provides insight that the components with identical $\upsilon_z$ can be connected to get a material which can fold and unfold freely (e.g., see Fig. 3F). Additionally, using this insight, we can create numerous configurations of metamaterials (see Appendix for more details). For the sheets made of tessellation of the same BCH$_n$ (e.g., Fig. 3A), $\upsilon_{e-e}$ is given by

$$(\upsilon_{WL})_{e-e} = -\frac{\varepsilon_L}{\varepsilon_W} = -\frac{dL/L}{dW/W} = -\tan^2\phi \frac{\kappa \cdot \lambda \cos\alpha - \cos^2\phi}{\kappa \cdot \lambda \cos\alpha + \cos^2\phi} \tag{7}$$

$$\text{in which, } \kappa = \frac{2n \cdot m_1}{m_1(n-1)+1} \quad \text{and} \quad \lambda = a/b \tag{8}$$

in which $n=2$ (notice that $n=1$ gives the relation for the Miura-ori sheet). Considering the end-to-end dimension, for a unit cell of BCH$_2$ ($m_1=1$), the $\upsilon_{e-e}$ is similar to that of a Miura-ori unit cell (Fig. 4C) and is given by

$$(\upsilon_{WL})_{e-e} = -\tan^2\phi \frac{2\lambda\cos\alpha - \cos^2\phi}{2\lambda\cos\alpha + \cos^2\phi} \tag{9}$$

Therefore, unlike $\upsilon_z$ which is always negative (Fig. 4B), $\upsilon_{e-e}$ can be positive for some geometric ranges (Fig. 4 C and D). Moreover, $\upsilon_z$ is only a function of the angle $\phi$, but $\upsilon_{e-e}$ can be dependent on other geometric parameters, i.e., the geometry of the facets ($a, b, \alpha$), tessellations ($n, m_1$) and the angle $\phi$. Notice that Poisson's ratio considering the end-to-end dimensions can be positive even for a Miura-ori unit cell (Fig. 4C). Also, note that the shift from negative to positive Poisson's ratio in Miura-ori is only the effect of considering the tail [31],, and the difference between two Poisson's ratios (i.e., $\upsilon_z$ and $\upsilon_{e-e}$) vanishes as the length of the Miura-ori sheet approaches infinity. However, note that for the BCH patterns, the transition to positive Poisson's ratio is mainly the effect of the holes in the sheets, and unlike the Miura-ori, even for a BCH sheet with an infinite configuration, the difference between two the approaches (i.e., $\upsilon_z$ and $\upsilon_{e-e}$) does not disappear (see Fig. 5). In this regard, Fig. 5 presents Poisson's ratio of a repeating unit cell of BCH$_2$ pattern within an infinite tessellation which corresponds to the following expression.

$$(\upsilon_\infty)_{e-e} = -\tan^2\phi \frac{4\lambda\cos\alpha - \cos^2\phi}{4\lambda\cos\alpha + \cos^2\phi} \tag{10}$$

From equation (10), $(\upsilon_\infty)_{e-e}$ for the BCH$_2$ sheet is positive, if $4\lambda\cos\alpha < \cos^2\phi$. The value is negative, if $4\lambda\cos\alpha > \cos^2\phi$.

Analogous to the Miura-ori sheet [1], similar BCH sheets possessing the same $\upsilon_z$ can be designed to be stacked and attached together along joining fold lines to form cellular folded metamaterials with capability of folding freely (Fig. 3 G and H). Note that the BCH sheets tailored for the stacking have identical $\upsilon_{e-e}$ as well (see Appendix for more details).

Considering that the facets are rigid and are connected via elastic rotational springs along the fold lines, we obtain the planar stretching stiffness of BCH$_2$, in both *x* and *y* directions (Fig. A5) and compare the results with their corresponding values for the Miura-ori cell. From Fig. 6, we infer that, depending on the geometry, and considering the same amount of material (compare Fig. 2A with Fig. A4), BCH$_2$ can be more or less stiff than its corresponding Miura-ori cell in the *x* and *y* directions.

Similarly to Miura-ori pattern, simple experimental observations show that these folded sheets exhibit anticlastic (saddle shape) curvature under bending (see Fig. 7A, Fig. A6A to Fig. A8A), which is an adopted curvature by conventional materials with positive out-of-plane Poisson's ratio [28]. This positional semi-auxetic behavior has been observed in the 'anti-trichiral' honeycomb [32] and auxetic composite laminates [33], as well as in other patterns of folded sheets made of conventional materials [1, 3, 27].

We investigated the effect of the geometry and material properties on the global behavior of the folded sheets, using the bar-framework numerical approach described by Schenk and Guest [1]. By considering the bending stiffness of the facets and fold lines ($K_{facet}$ and $K_{fold}$, respectively), we studied the modal responses of the folded shells by changing the ratio of $K_{facet}$ to $K_{fold}$. For the BCH$_2$ pattern, shown in Fig. 7, similarly to a regular Miura-ori sheet [27], twisting and bending modes are the predominant behavior of the pattern over a range of $K_{facet}/K_{fold}$ and associated geometries (Fig. 7, B and C). Furthermore, the saddle-shaped bending mode obtained from eigenvalue analysis of the patterns further confirms that the Poisson's ratio under bending is positive [28]. The results show that for large values of $K_{facet}/K_{fold}$, the first softest eigen-mode represents a rigid origami behavior (Fig. 7D). The results of the stiffness analysis for several other patterns from the class of metamaterials show similar behavior (see Fig. A6 to Fig. A8).

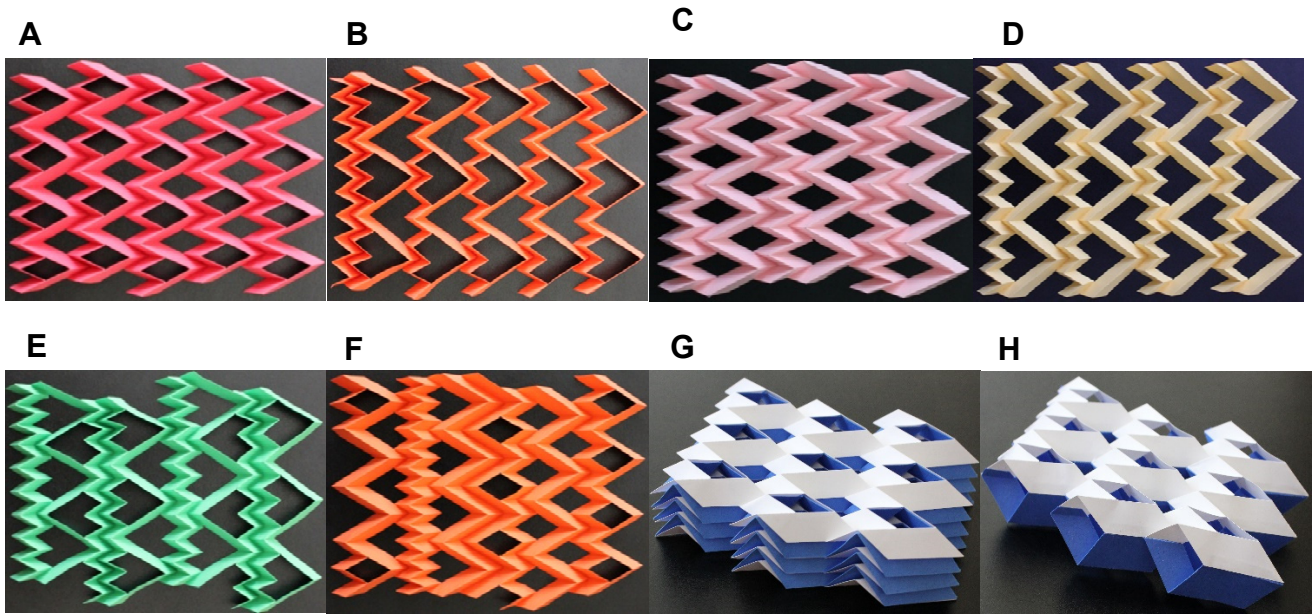

**Fig. 3. Sample patterns including BCH$_n$ and cellular folded metamaterials.** (**A**) A sheet of BCH$_2$. (**B**) A sheet of BCH$_3$. Adding one layer of small parallelograms to the first row reduces the DOFs of the system to 1 for rigid origami behavior. (**C**) Combination of BCH$_2$ and layers of large and small parallelograms with the same geometries as the ones used in the BCH$_2$. (**D**) Combination of BCH$_3$ and layers of large and small parallelograms with the same geometries as the ones used in the BCH$_3$. (**E**) Sheet of BCH$_3$ and layers of small parallelograms with the same geometries as the ones used in the BCH$_3$. (**F**) A sheet composed of various BCH$_n$ and Miura-ori cells with the same angle $\phi$. (**G**) A stacked cellular metamaterial made from 7 layers of folded sheets of BCH$_2$ with two different geometries. (**H**) Cellular metamaterial made from 2 layers of 3x3 sheets of BCH$_2$ with different heights tailored for the stacking purpose, and bonded along the joining fold lines. The material is flat-foldable in one direction.

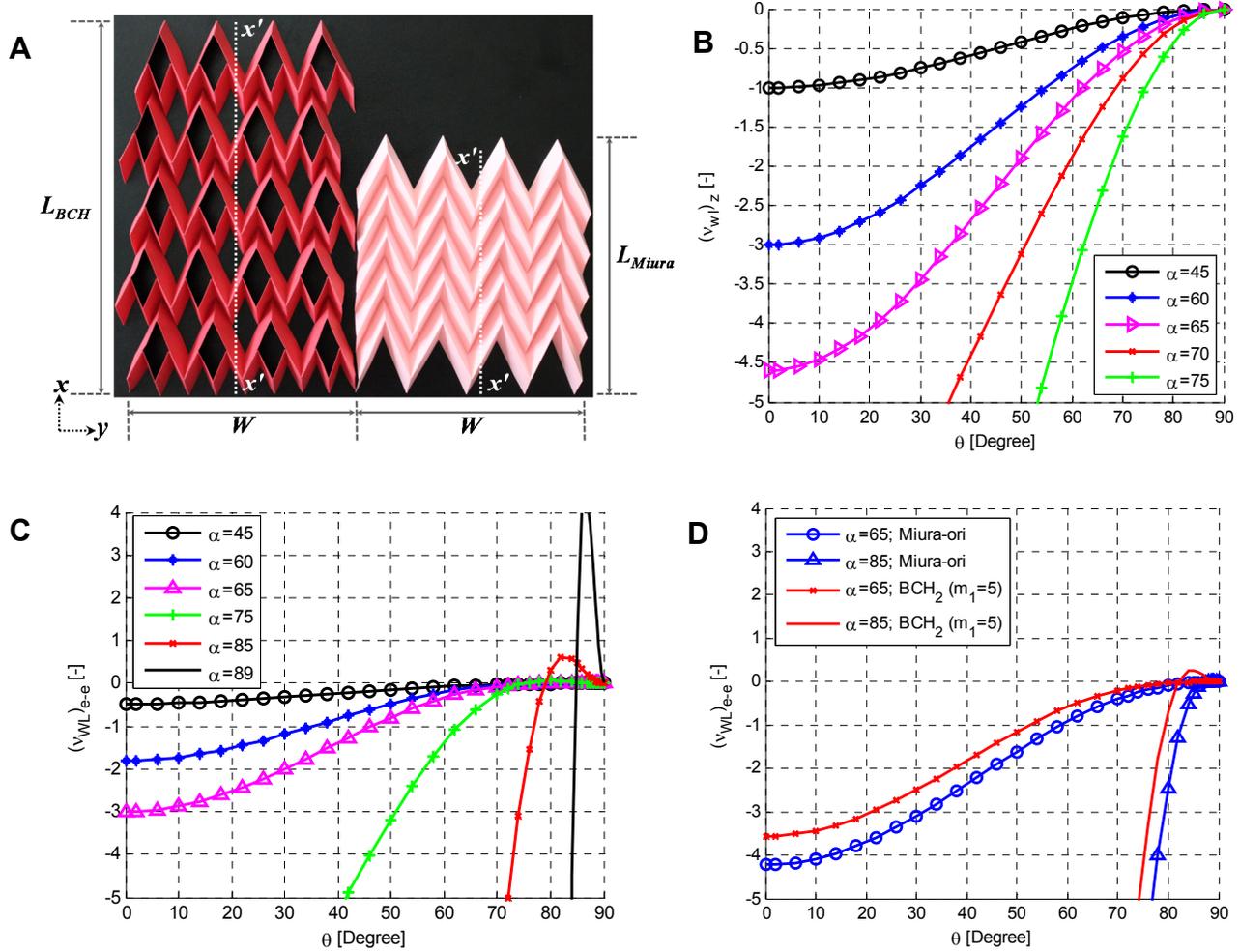

**Fig. 4. In-plane Poisson's ratios of finite configurations of the metamaterials.** (**A**) 5x4 ($m_1$=5 and $m_2$=4) BCH$_2$ sheet (left image) and its corresponding Miura-ori sheet (right image) with the same geometry, and the same amount of material. Projected lengths of the zigzag strips along *x'-x'* line parallel to the *x*-axis is used to obtain $\upsilon_z$ and $L$ is used to obtain $\upsilon_{e-e}$. Both sheets have identical $\upsilon_z$, but they have different $\upsilon_{e-e}$. (**B**) In-plane kinematics ($\upsilon_z$) for the class of metamaterials. (**C**) In-plane Poisson's ratio considering the end-to-end dimensions ($\upsilon_{e-e}$) for a single unit cell of Miura-ori and BCH$_2$ patterns with *a=b*. (**D**) In-plane Poisson's ratio considering the end-to-end dimensions ($\upsilon_{e-e}$) for sheets of Miura-ori and BCH$_2$ with $m_1$=5 and *a=b*.

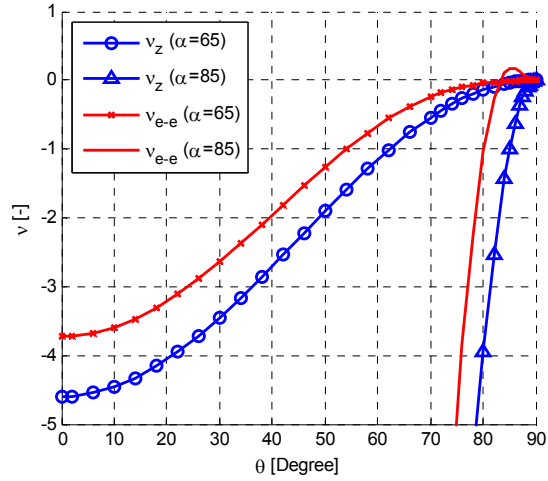

**Fig. 5. Poisson's ratio of BCH$_2$ sheet for an infinite configuration.** Poisson's ratio obtained by considering the projected length of the zigzag strips, $\upsilon_z$, versus Poisson's ratio considering the end-to-end dimensions of the sheet when the sheet size approaches infinity, $\upsilon_{e-e}$ (a=b and $m_1 \to \infty$). The latter is equivalent to the Poisson's ratio of a repeating unit cell of BCH$_2$ within an infinite tessellation. The figure shows that, contrary to the Miura-ori, the transition towards positive Poisson's ratio is present for an infinite configuration of the BCH$_2$ sheet.

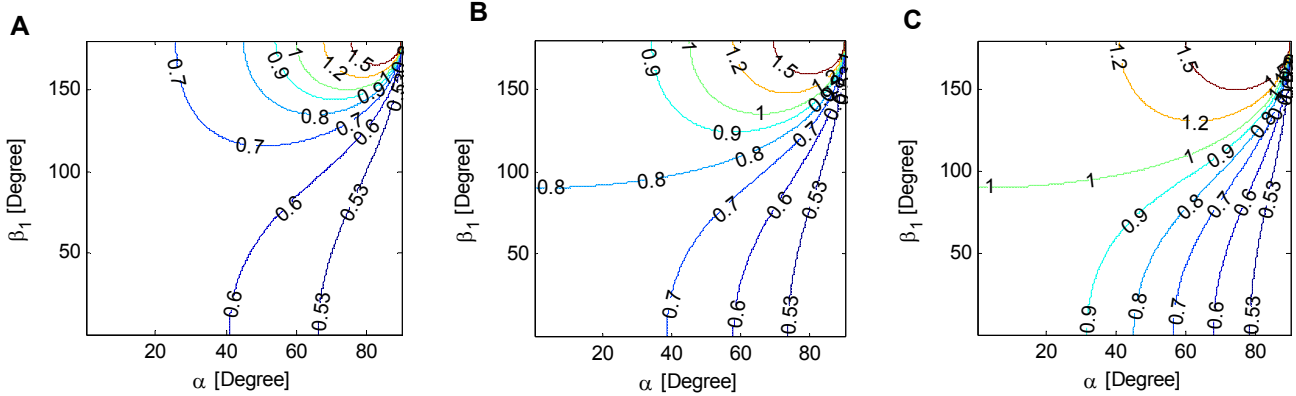

**Fig. 6. Ratio of in-plane stiffness of Miura-ori cell to that of the BCH$_2$ in the *x* and *y* directions.** The results show that depending on the geometry and considering the same amount of material, BCH$_2$ can be more or less stiff than its corresponding Miura-ori cell in the *x* and *y* directions. (**A**) *a/b=2*. (**B**) *a/b=1*. (**C**) *a/b=1/2*.

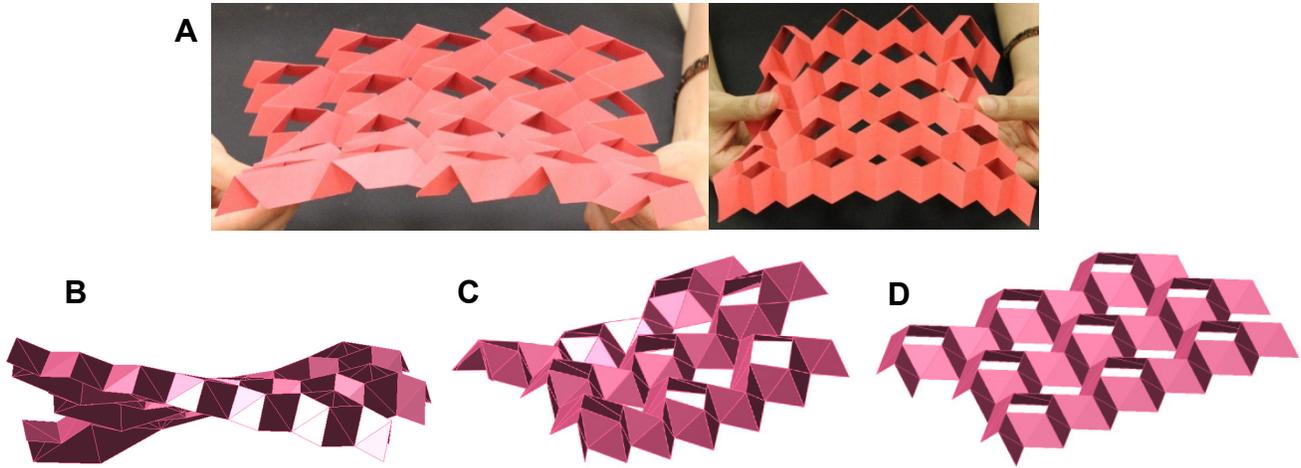

**Fig. 7. Behavior of the sheet of the $BCH_2$ under bending and the results of the eigen-value analysis of a 3 by 3 $BCH_2$ pattern.** (**A**) A sheet of $BCH_2$ deforms into a saddle-shaped under bending (i.e., a typical behavior seen in materials with a positive out-of-plane Poisson's ratio). (**B**) Twisting, (**C**) saddle-shaped and (**D**) rigid origami behavior (planar mechanism) of a 3 by 3 pattern of $BCH_2$ ($a=1$; $b=2$; $\alpha = 60°$). Twisting and saddle-shaped deformations are the softest modes observed for a wide range of material properties and geometries. For large values of $K_{facet}/K_{fold}$ the rigid origami behavior (planar mechanism) is the softest deformation mode of the sheets.

## Discussion

In the present research, we have employed the concept of the in-plane Poisson's ratio in two different contexts: Firstly, to describe the kinematics and to create a class of one-DOF zigzag-base mechanical metamaterials. To address this point, the Poisson's ratio is obtained by considering the projected lengths of the zigzag strips, $\upsilon_z$, and the value is always equal to $(-\tan^2 \phi)$. The Poisson's ratio obtained in this way is an inherent property of the class of one-DOF zigzag-base folded sheets, and is related to the foldability of the class of the metamaterials. Hence, the concept is insightful to create novel zigzag-base foldable materials. Note that the point has been used in the literature [1] to describe the stacking of the Miura-ori, by associating the value (i.e., $-\tan^2 \phi$) to the Poisson's ratio of the Miura-ori sheet [1]. However, in the present work, by explicitly associating the value to that of a one-DOF zigzag strip (Fig. 1), we have created the BCH patterns containing various scales of zigzag strips. Accordingly, the present study, by introducing the BCH patterns extends the kinematics of the Miura-ori to that of a class of one-DOF zigzag-base folded sheet metamaterials. In other words, our work states that all one-DOF zigzag-base folded metamaterials shown in Fig. 3 have identical kinematics, if the angle $\phi$ is the same. Secondly, to study the size change of a folded metamaterial introduced in this work. In this regard, the Poisson's ratio is obtained by considering the end-to-end dimensions of the sheet, $\upsilon_{e-e}$. Note that this definition captures the size change of a finite sheet (Fig. 4) and that of a repeating unit cell within an infinite configuration for a regular sheet (e.g., regular $BCH_2$ - see Fig. 5). Moreover, it is applicable for irregular sheets such as the one shown in Fig. 3F.

Because recent literature on the topic had different arguments regarding Poisson's ratio evaluation [1, 3, 31], this study, by introducing a class of zigzag-base folded sheet materials with holes, clarifies the issue and unifies the concepts. In this regard, note that for the Miura-ori sheet, the Poisson's ratio of a repeating unit cell within an infinite tessellation is equal to $v_z$. Hence, the value given in [1, 3] presents both the kinematics of the Miura-ori sheet and the size change of a repeating unit cell within an infinite configuration. Thus, considering the end-to-end dimensions in a finite Miura-ori sheet is simply to capture the edge effect. However, for the $BCH_2$ pattern, the Poisson's ratio of a repeating unit cell within an infinite configuration is not equal to $v_z$, and it assumes both negative and positive Poisson's ratios due to the presence of the holes in the pattern (Fig. 5). Therefore, considering the end-to-end configuration for the $BCH_2$ pattern is mainly to capture the effect of the holes in the Poisson's ratio.

We have also shown that the $BCH_n$ and combined patterns, introduced in this work, possess metamaterial properties arising from their tunable geometrical configurations. An appealing feature of these patterns is that they display similar properties to those of the Miura-ori, however, presence of the different scales of zigzag strips in the structure of the patterns, as well as existence of holes, make the $BCH_n$ patterns unique (e.g., see Fig. 4A). In addition, the fact that the $BCH_n$ mechanical properties differ from those of the Miura-ori (e.g., see Fig. 4D, Fig. 5 and Fig. 6) offer avenues to explore alternative materials and structures based on these patterns for a specific performance/application of the Miura-ori pattern on which there is a surge of research interest. On the other hand, present technology requires lighter and more customizable structures and materials. Combining cellular $BCH_n$ patterns with Miura-ori provides an augmented design space for tailored engineering design of materials and structures. Consequently, availability of large design motifs can be advantageous, for instance, in dynamic architectural façades where the place of the holes in the patterns can be controlled by combining Miura-ori with $BCH_n$ to either allow light in the interior of the building or to promote shading when desirable.

In Summary, the remarkable properties of the patterns (specifically $BCH_2$), such as rigid-foldability, flat-foldability, and possessing single DOF, as well as numerous possible combinations of the patterns make them potentially suited for a broad range of applications including kinetic and deployable structures (e.g., solar sails [24]), light cellular foldcore sandwich panels [25, 26], 3D tunable folded cellular metamaterials [1, 4, 34], energy absorbing devices [35], foldable robots [36] and auxetic materials [28, 29]. In all these applications, scalability is a major feature of the $BCH_n$ or other combined patterns (due to their inherent geometric properties).

[28] A. Alderson and K. L. Alderson, Auxetic Materials. *Proc. IMechE J. Aerospace Eng.* **221**, 565-575 (2007).

[29] R. Lakes, Foam structures with a negative Poisson's ratio. *Science* **235**, 1038-1040 (1987).

[30] T. C. T. Ting, T. Chen, Poisson's ratio for anisotropic elastic materials can have no bounds. *Q. J. Mech. Appl. Math* **58**, 73-82 (2005).

[31] C. Lv, D. Krishnaraju, G. Konjevod, H. Yu, H. Jiang. Origami based Mechanical Metamaterials. *Scientific Reports* **4**, 5979 (2014).

[32] A. Alderson, K. L. Alderson, G. Chirima, N. Ravirala, and K. M. Zied, The in-plane linear elastic constants and out-of-plane bending of 3-coordinated ligament and cylinder-ligament honeycombs. *Compos. Sci. Technol.* **70**, 1034-1041 (2010).

[33] T. C. Lim, On simultaneous positive and negative Poisson's ratio laminates. *Phys. Status Solidi B* **244**, 910-918 (2007).

[34] K. C. Cheung, T. Tachi, S. Calisch, K. Miura, Origami interleaved tube cellular materials. *Smart Mater. Struct.* **23**, 094012 (2014).

[35] M. Schenk, S. D. Guest and G. J. McShane, Novel stacked folded cores for blast-resistant sandwich beams. *IJSS* **51**, 4196-4214 (2014).

[36] S. Felton, M. Tolley, E. Demaine, D. Rus, R. Wood, A method for building self-folding machines. *Science* **345**, 644 (2014).

[37] W. McGuire, R.H. Gallagher, R.D. Ziemian, *Matrix Structural Analysis* (Wiley, New York, ed. 2, 2000).
Maryam Eidini, Glaucio H. Paulino     Main Text   **13**/**13**

# Appendix

## 1- Geometry, pattern tessellation and combination

The geometry of $BCH_n$ (Basic unit Cell with Hole) is described in the main text and is shown in Fig. 2A. By combining $BCH_n$ with row/rows of small and/or large parallelograms with the same angle $\phi$, we can obtain numerous unit cells. A few configurations are presented in Fig. A1. The tessellations or/and combinations of tessellation of these cells having the same angle $\phi$ can result in a new metamaterial. For example, the patterns shown in Fig. 3 C and D are obtained by tessellations of the unit cells presented in Fig. A1D.

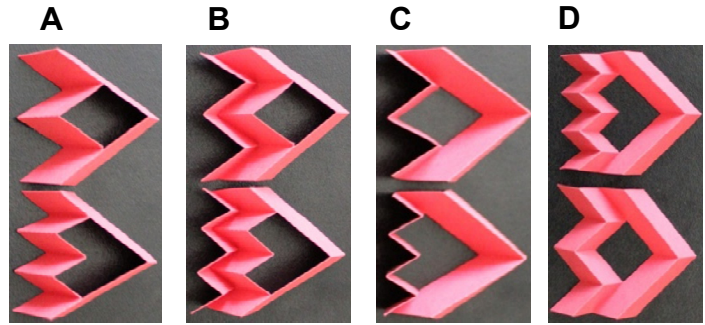

**Fig. A1. $BCH_2$ and $BCH_3$ and their combinations with row/rows of small and/or large parallelograms.** (**A**) A $BCH_2$ and a $BCH_3$. (**B**) A $BCH_2$ and a $BCH_3$ combined with a row of small parallelograms with the same geometry as the one used in the corresponding BCH. (**C**) A $BCH_2$ and a $BCH_3$ combined with a row of large parallelograms with the same geometry as the one used in their corresponding BCH. (**D**) A $BCH_2$ and a $BCH_3$ combined with rows of small and large parallelograms with the same geometry as the one used in their corresponding BCH.

## 2- Number of degrees of freedom of the patterns

For the case of $n=2$, i.e., for the unit cell of $BCH_2$ pattern shown in Fig. 2A, and for a given geometry of the facet, the geometry of the unit cell implies that it can be defined based on only one fold angle (i.e., similarly to Miura-ori (*1*), we can write the relations between all degrees of freedom (DOFs) based on only one fold angle, e.g., the angle $\phi$). Hence, the unit cell of $BCH_2$ has only one DOF. On the other hand, implicit formation of the structure of the Miura-ori unit cell with one DOF mechanism between two adjoining unit cells, as shown in Fig. A2A, imposes the whole pattern to have only one DOF. The conclusion is further verified using numerical calculation of the number of DOFs as described in Section (6-1).

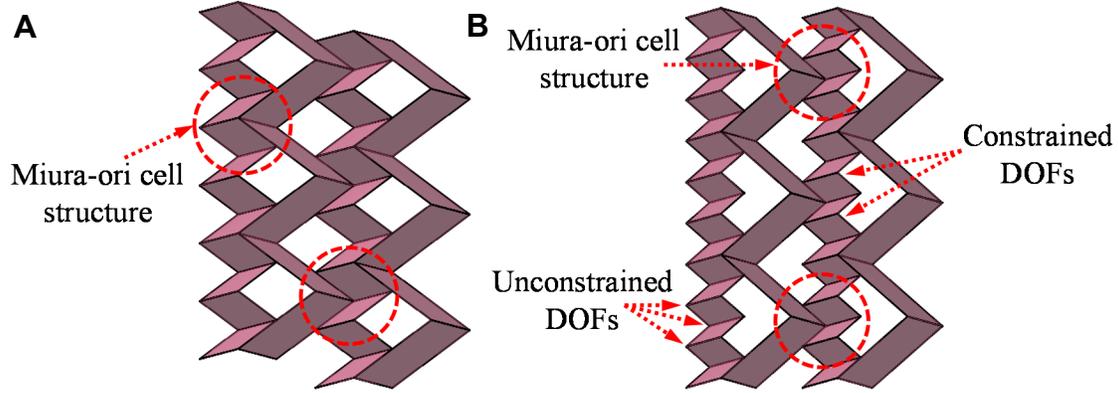

**Fig. A2. Constrained DOFs by implicit formation of the structure of the Miura-ori unit cell between adjoining unit cells of BCH$_2$ and BCH$_3$ in the pattern.** (**A**) The Miura-ori unit cell structure formed implicitly in the tessellation makes the whole BCH$_2$ pattern fold with one-DOF planar mechanism. (**B**) In the symmetric tessellation of identical BCH$_3$, except for the small parallelogram facets of the first row, all other independent DOFs in the unit cell of BCH$_3$ are constrained by the structure of the Miura-ori cell formed between two adjoining unit cells.

In general, from the numerical model described in Section (6-1) and for rigid origami behavior, the unit cell of BCH$_n$ (Fig. A1A) has 2$n$-3 DOFs. However, adding one row of small parallelogram facets to the BCH$_n$ (e.g., Fig. A1B), i.e., creating a complete row of Miura-ori unit cells with the small parallelogram facets, reduces the DOFs of the cell to 1 irrespective of the number of $n$. Hence, tessellation of the unit cells shown in Fig. A1B can create patterns with one DOF planar mechanism (e.g., Fig. 3E).

For the symmetric tessellation of BCH$_3$, as presented in Fig. A2B, except for the first row of small parallelogram facets, all other independent DOFs in the unit cell of BCH$_3$ are constrained by the implicit formation of the structure of Miura-ori cell between two adjoining unit cells. Hence, the pattern of BCH$_3$, shown in Fig. A2B, due to existence of unconstrained DOFs in the first row of small parallelogram facets, has more than one DOF. However, adding one row of small parallelogram facets and, accordingly, creating the row of Miura-ori cells with small parallelogram facts can reduce the DOF of the whole system to one (e.g., Fig. 3B).

## 3- In-plane stretching response of BCH$_n$ sheets

### 3-1- Poisson's ratio

A 5x4 sheet of BCH$_2$ along with its corresponding Miura-ori sheet containing the same geometry of facets and fold angle ($a$, $b$, $\alpha$ and $\phi$ are identical in both models) is shown in Fig. 4A. The Poisson's ratio $\upsilon_z$ for both sheets can be obtained from the following relation

$$\left(\upsilon_{w\ell}\right)_z = -\frac{\varepsilon_\ell}{\varepsilon_w} = -\frac{d\ell/\ell}{dw/w} \tag{1}$$

in which $\ell$ is the projected length of the zigzag strips in the **xy**-plane and parallel to the **x**-axis (*i.e.*, the projected lengths of the strips along any arbitrary lines of **x'-x'** in the **xy** plane and parallel to the **x**-axis intersecting a complete tessellation). Hence, for a sheet with $m_1$ rows, $\ell$ is equal to $m_1$ times the projected lengths of the strips in the unit cell shown in Fig. 2A. From Fig. 4A, the importance of considering the end-to-end dimensions to obtain the Poisson's ratio for folded sheets is more pronounced in sheets with holes, because both sheets have the same $\upsilon_z$ despite having different lengths along the **x**-axis. Another example, showing the relevance of the end-to-end dimensions to obtain Poisson's ratio of folded sheets is presented in Fig. A3, where two identical 2x2 Miura-ori sheets are shown. Moreover, 2 rows of small Mira-ori cells with equal $\upsilon_z$ are also attached to the left-hand sample as shown in the figure. Therefore, from the figure, considering the end-to-end dimensions to obtain Poisson's ratio in the left system is obvious. For the limit cases of very large number of small cells, as well as very small length of *a* for small cells, *i.e.*, $n \to \infty$ and $a_s \to 0$ (See Fig. A4 for the geometry of Miura-ori cell), the rows of small Miura-ori cells approach the lines defining the end-to-end dimension for the 2x2 Miura-ori sheet shown on the right-hand image.

From the expression (7) in the main text, when $\kappa \cdot \lambda \cos\alpha < \cos^2\phi$, $\upsilon_{e-e}$ is positive and for $\kappa \cdot \lambda \cos\alpha > \cos^2\phi$, $\upsilon_{e-e}$ is negative.

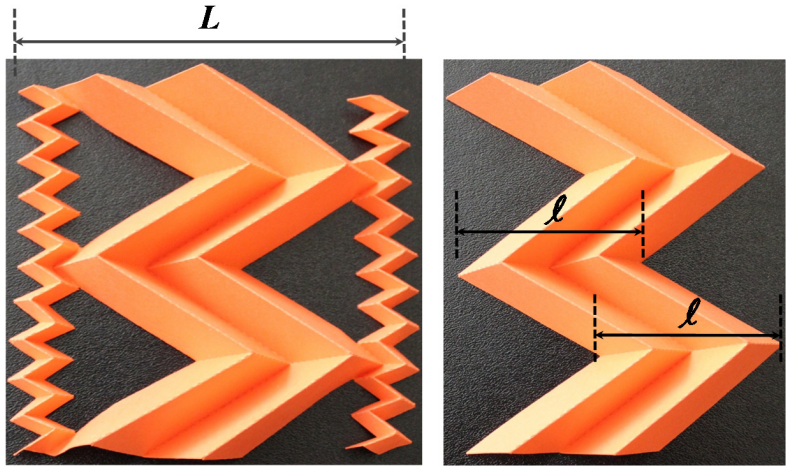

**Fig. A3. Concept of Poisson's ratio considering the end-to-end dimensions.** Figure shows two identical 2x2 Miura-ori tessellations. The 2 rows of small Mira-ori cells with the same $\upsilon_z$ as that of the 2x2 sheet are attached to the left sample. Length *b* of the small cells are 1/5 of that of the large cells (*i.e.*, the number of small cells per each large cell is 5 (*n*=5)).

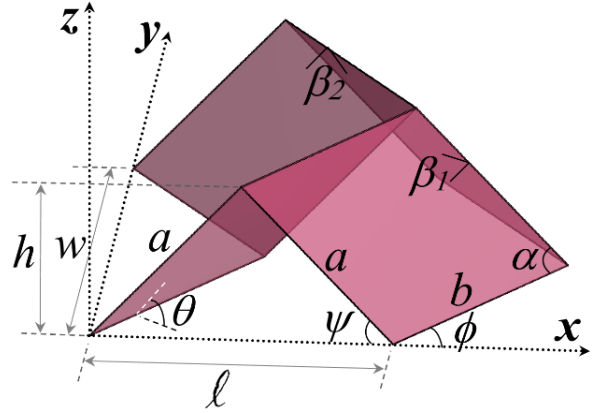

**Fig. A4. Geometry of Miura-ori cell.**

For a Miura-ori ($n=1$) sheet, we have

$$(\upsilon_{WL})_{e-e} = -\tan^2\phi \frac{2m_1\lambda\cos\alpha - \cos^2\phi}{2m_1\lambda\cos\alpha + \cos^2\phi} \qquad (2)$$

Hence, for a Miura-ori sheet, if $m_1 \to \infty$, then $\upsilon_{e-e}$ approaches $\upsilon_z$ (i.e., $-\tan^2\phi$). Notice that, from the above relation, even a Miura-ori unit cell (i.e., $m_1 = 1$) can have positive Poisson's ratio for some ranges.

For a sheet of BCH$_2$, we have

$$(\upsilon_{WL})_{e-e} = -\tan^2\phi \frac{\frac{4m_1}{m_1+1}\lambda\cos\alpha - \cos^2\phi}{\frac{4m_1}{m_1+1}\lambda\cos\alpha + \cos^2\phi} \qquad (3)$$

Note that from the above relation, for a sheet of BCH$_2$, while comparing $\upsilon_z$ with $\upsilon_{e-e}$, the shift towards positive Poisson's ratio in $\upsilon_{e-e}$ is mainly the effect of the holes, and thus the difference between $\upsilon_z$ and $\upsilon_{e-e}$ does not disappear (see Fig. 5) when the length of the sheet approaches infinity ($m_1 \to \infty$).

### 3-2- Stretching stiffness
### 3-2-1- BCH$_2$

In this section, we derived the in-plane stiffness of the BCH$_2$ in the *x* and *y* directions. For this purpose, an alternative parameterization for BCH$_2$ (the unit cell is shown in Fig.1, A and B) based on the dihedral angles between the rigid facets is used, which is similar to the equations of reference [3] for Miura-ori cell. To better compare the results, we kept the same symbols as those given for Miura-ori [3], provided that they are consistent with the symbols used in this work. Therefore,

$$\ell = 2a\zeta \qquad w = 2b\xi \qquad h = a\zeta\tan\alpha\cos(\beta_1/2) \qquad (4)$$

in which,

$$\xi = \sin\alpha\sin(\beta_1/2) \quad \text{and} \quad \zeta = \cos\alpha\left(1-\xi^2\right)^{-1/2} \qquad (5)$$

The potential energy of a BCH$_2$ (see Fig.1, A and B), subjected to uniaxial force in the *x* direction, can be obtained from

$$H = U + \Omega \qquad (6)$$

in which, $U$ and $\Omega$ are elastic energy and potential of the applied load, respectively that are given by

$$U = \frac{1}{2}k\left(4a(\beta_1-\beta_{1_0})^2 + b(\beta_2-\beta_{2_0})^2\right) \qquad (7)$$

$$\Omega = -\int_{\beta_{1_0}}^{\beta_1} f_x \frac{d\ell}{d\beta_1'} d\beta_1' \qquad (8)$$

where $k$ is the rotational hinge spring constant per unit length ($k=K_{fold}$). Setting the condition $\partial H/\partial\beta_1 = 0$, the external force at equilibrium $f_x$ can be obtained from

$$f_x = \frac{dU/d\beta_1}{d\ell/d\beta_1} = \frac{2k}{\cos\alpha\sin^2\alpha}\left(\frac{4(\beta_1-\beta_{1_0})\left(1-\xi^2\right)^{3/2} + \frac{b}{a}\cdot(\beta_2-\beta_{2_0})\cos\alpha\left(1-\xi^2\right)^{1/2}}{\sin\beta_1}\right) \qquad (9)$$

Notice that $K_x$ is obtained from the following expression

$$K_x(\alpha,\beta_{1_0}) = \left.\frac{df_x}{d\ell}\right|_{\beta_{1_0}} = \left.\frac{df_x}{d\beta_1}\cdot\frac{d\beta_1}{d\ell}\right|_{\beta_{1_0}} \qquad (10)$$

$$K_x(\alpha,\beta_{1_0}) = \left(2k\cdot\frac{4a(1-\xi_0^2)^2 + b\cos^2\alpha}{a(1-\xi_0^2)^{1/2}\cos\alpha\sin^2\alpha\sin\beta_{1_0}}\right)\frac{8a^2\cos\alpha}{\ell^3\sqrt{1-\frac{4a^2\cos^2\alpha}{\ell^2}}\sqrt{\frac{4a^2}{\ell^2}-1}} \qquad (11)$$

where, $\xi_0 = \xi(\alpha,\beta_{1_0})$. The contour plot of the stiffness ratio in the *x* direction ($K_x/k$) is shown in Fig. A5A for a unit cell with $a=b=1$, in terms of the facet angle $\alpha$ and fold angle $\beta_1$.

Similarly, for the *y* direction:

$$f_y = \frac{dU/d\beta_1}{dw/d\beta_1} = \frac{k}{b\sin\alpha} \cdot \frac{4a(\beta_1 - \beta_{1_0}) + b(\beta_2 - \beta_{2_0})\left(\frac{\cos\alpha}{1-\xi^2}\right)}{\cos(\beta_1/2)} \quad (12)$$

$$K_y(\alpha, \beta_{1_0}) = \left.\frac{df_y}{dw}\right|_{\beta_{1_0}} = \left.\frac{df_y}{d\beta_1} \cdot \frac{d\beta_1}{dw}\right|_{\beta_{1_0}} \quad (13)$$

$$K_y(\alpha, \beta_{1_0}) = \frac{k}{b}\left(\frac{4a(1-\xi_0^2)^2 + b\cos^2\alpha}{(1-\xi_0^2)^2 \sin\alpha \cos(\beta_{1_0}/2)}\right)\left(\frac{2}{\sqrt{4b^2\sin^2\alpha - w^2}}\right) \quad (14)$$

The contour plot of the stiffness in the *y* direction is shown in Fig. A5B for a unit cell with $a=b=1$ in terms of the facet and fold angles, $\alpha$ and $\beta_1$, respectively.

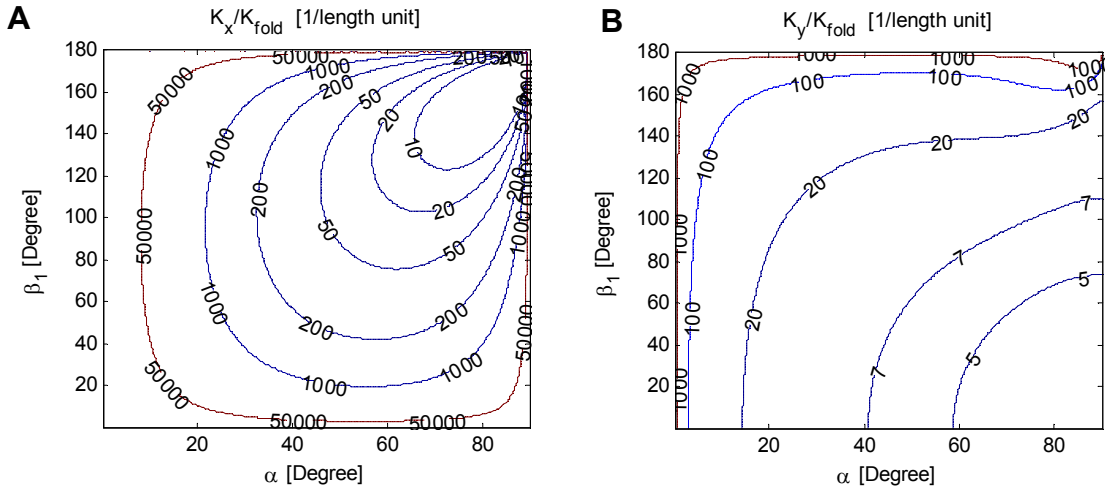

**Fig. A5. In-plane stiffness for the BCH2 with *a=b=1*.** (**A**) $K_x/k$. (**B**) $K_y/k$.

### 3-2-2- Classical Miura-ori

The in-plane stretching stiffness in the *x* direction for Miura-ori cell is obtained by equation (10), which results in

$$K_x(\alpha, \beta_{1_0}) = \left(4k\frac{a(1-\xi_0^2)^2 + b\cos^2\alpha}{a(1-\xi_0^2)^{1/2}\cos\alpha\sin^2\alpha\sin\beta_{1_0}}\right)\left(\frac{8a^2\cos\alpha}{\ell^3\sqrt{1-\frac{4a^2\cos^2\alpha}{\ell^2}}\sqrt{\frac{4a^2}{\ell^2}-1}}\right) \quad (15)$$

The second term within brackets is missing in Ref. [3]. Similarly, the stretching stiffness in the *y* direction is obtained by equation (13), which results in

$$K_y(\alpha, \beta_{l_0}) = \left(2k \frac{a(1-\xi_0^2)^2 + b\cos^2\alpha}{b(1-\xi_0^2)^2 \sin\alpha \cos(\beta_{l_0}/2)}\right)\left(\frac{2}{\sqrt{4b^2\sin^2\alpha - w^2}}\right) \qquad (16)$$

Notice that $K_x/k$ and $K_y/k$ are not dimensionless, and thus $K_x$ and $K_y$ have dimension of in-plane stretching stiffness [37].

### 3-2-3- In-plane Stiffness of BCH2 compared with its corresponding Miura-ori cell

Figure 5 shows the ratio of the in-plane stiffness of the Miura-ori cell in the *x* and *y* directions to those of the BCH2, for various ratios of *a/b*. The ratio is equal for both *x* and *y* directions because only the numerators of the first term in the planar stiffness relations change from BCH2 to Miura-ori cell, and the numerators are equal in both planar rigidities of *x* and *y* directions for a specific unit cell.

## 4- BCH$_n$-based Cellular folded metamaterial

Similar sheets with different heights, while possessing the same $\upsilon_z$, can be stacked and attached together along joining fold lines to make a cellular folded metamaterials (see Fig. 3 G and H). The fold geometry can change from layer to layer, but assuming stacking of the patterns using 2 layers of A and B [1], and by equating the external dimensions as well as $\upsilon_z$ for both layers of A and B, respectively, we have

$$b_B = b_A \qquad a_B = a_A \frac{\cos\alpha_A}{\cos\alpha_B} \qquad (17)$$

$$\theta_B = \arccos\left(\cos\theta_A \frac{\tan\alpha_A}{\tan\alpha_B}\right) \qquad (18)$$

where the geometry of the layer B can be obtained based on that of the layer A. It is worth noting that meeting the above equations for the stacking of the layers results in the sheets possessing identical $\upsilon_{e-e}$ as well as shown in the following.

$$(\upsilon_{e-e})_A = -\tan^2\phi \frac{\kappa \frac{a_A}{b_A}\cos\alpha_A - \cos^2\phi}{\kappa \frac{a_A}{b_A}\cos\alpha_A + \cos^2\phi} = -\tan^2\phi \frac{\kappa \frac{a_B}{b_B}\cos\alpha_B - \cos^2\phi}{\kappa \frac{a_B}{b_B}\cos\alpha_B + \cos^2\phi} = (\upsilon_{e-e})_B \qquad (19)$$

This result further emphasizes the relevance of defining the end-to-end Poisson's ratio for the zigzag-base sheet metamaterials.

The angle $\theta$ for the stacked samples shown in Fig. 3G is positive for both sheets, *i.e.* $\theta_A, \theta_B \in [0, \pi/2]$. This form of stacking may be applied as impact absorbing devices [35].

Considering the angle $\theta$ for one alternating layer being negative, *i.e.* $\theta_A \in [-\pi/2, 0]$ results in a new metamaterial in which the layers can be connected along joining fold lines using adhesive (Fig. 3H). In this way of stacking, the heights for two successive layers can be identical.

## 5- Experimental Responses of the Patterns

### 5-1- In-plane behavior

Under in-plane extension, the patterns, for large geometric ranges (see the analytical model in the main text), exhibit negative Poisson's ratios. Simple in-plane experimental tests confirm that, for most geometric ranges, the patterns exhibit negative Poisson's ratio.

### 5-2- Out-of-plane behavior

Under bending, this class of patterns folded from various types of papers exhibit anticlastic (saddle-shaped) curvature (see Fig. 7A, Fig. A6A, Fig. A7A, and Fig. A8A) which is an adopted curvature by conventional materials with positive Poisson's ratio (25).

## 6- Numerical investigation of patterns behavior

To capture the effect of geometry and material properties on global behavior of a folded shell system, a stiffness analysis can be carried out and the structure can be simulated using a Finite Element Analysis (FEA). Depending on the application, either a constrained bar-framework origami modeling approach [1, 27] or a modeling scheme using nonlinear shell elements can be used at this stage. To model folded shell structures, we used the pin-jointed bar framework approach proposed by Schenk and Guest [1, 27]. In this modeling scheme, fold lines and vertices are modeled with bars and frictionless joints, respectively. To stabilize each facet and to model the bending of the facets for stiffness analysis, the facets are triangulated, and additional members are added to each facet. Placing of additional members is based on observations from physical models, stabilization of the facets, and energetic consideration in facet bending [1]. This model accounts for bending of the facets and the effect of out-of-plane kinematics of the sheets, and therefore is not restricted to rigid origami.

By varying the ratio of the bending stiffness of the facets and fold lines ($K_{facet}/K_{fold}$), we performed stiffness analysis for 3x3 patterns of BCH$_2$ and BCH$_3$, and 2x3 patterns of the unit cells shown in Fig. A1D with *α=60* degrees, *a=1* and *b=2*. The stiffness analysis of the patterns reveals that twisting and bending modes are predominant behavior of the patterns over a large range of $K_{facet}/K_{fold}$ and geometries (Fig. A6-Fig. A8, A and B). The modal shapes corresponding to the lowest eigen-value of the sheets show that for large values of $K_{facet}/K_{fold}$, the first softest deformation mode represents a rigid origami behavior (Fig. A6-Fig. A8, C).

### 6-1- Numerical calculation of the number of DOFs of the patterns

In the bar-framework analysis, *compatibility* equation is to relate the nodal displacements ***d*** to the bar extensions ***e*** via compatibility matrix ***C*** as follows

$$Cd = e \tag{20}$$

From the above equation, the nullspace of the compatibility matrix provides the solution in which the bars do not extend. To model rigid origami behavior, we need to add an angular constraint to the compatibility matrix whose nullspace can provide the nodal displacements $d$ for which the facets do not bend either. The angular constraint can be written in terms of the dihedral fold angles between triangulated facets connected by added fold lines [1, 27]. Hence,

$$J_{facet}\, d = d\rho \tag{21}$$

where $\rho$ is the dihedral fold angle between two adjoining triangulated facets and $J_{facet}$ is the Jacobian of the angular constraint considered for the triangulated facets intersecting by added fold lines. Therefore, the augmented compatibility matrix is as follows

$$\bar{C} = \begin{bmatrix} C \\ J_{facet} \end{bmatrix} \tag{22}$$

Accordingly, the number of internal infinitesimal mechanisms (*i.e.*, the number of independent DOFs) can be obtained from the expression

$$m = 3j - rank(\bar{C}) - 6 \tag{23}$$

in which $j$ is the number of joints (*i.e.*, the number of vertices). In the above relation, the 6 DOFs related to the rigid-body motions of 3D structures are excluded.

We used the above relation to obtain the number of DOFs for the patterns considering rigid origami behavior. The results are justified based on the geometry of the patterns as well as existence of the implicit formation of the structure of the Miura-ori cells with one-DOF mechanism as described in Section (2).

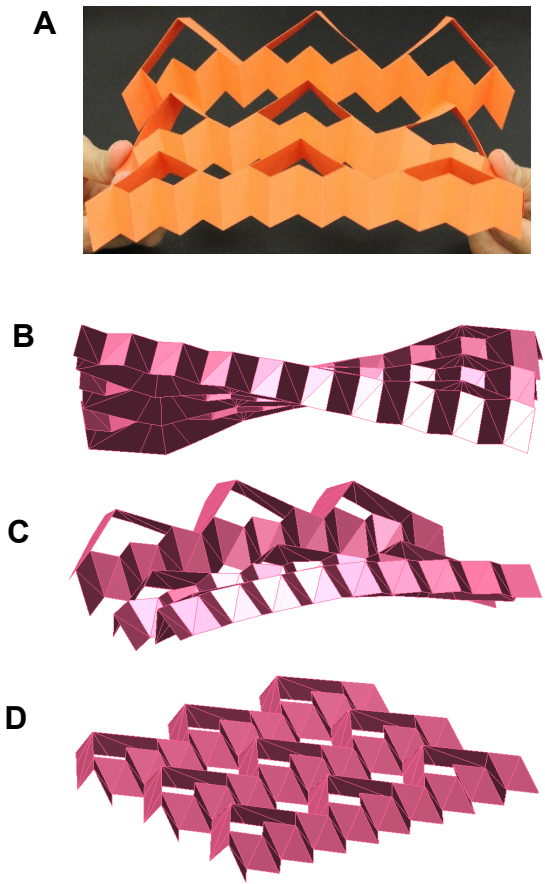

**Fig. A6. Behavior of the sheet of BCH$_3$ under bending and the results of eigen-value analysis of a 3 by 3 pattern of BCH$_3$.** (**A**) Sheet of BCH$_3$ deforms into a saddle-shaped under bending which is typical behavior for materials having a positive Poisson's ratio. (**B**) Twisting, (**C**) saddle-shaped and (**D**) rigid origami behavior (planar mechanism) of a 3 by 3 pattern of BCH$_3$ (with *a=1*; *b=2*; $\alpha = 60°$ ).

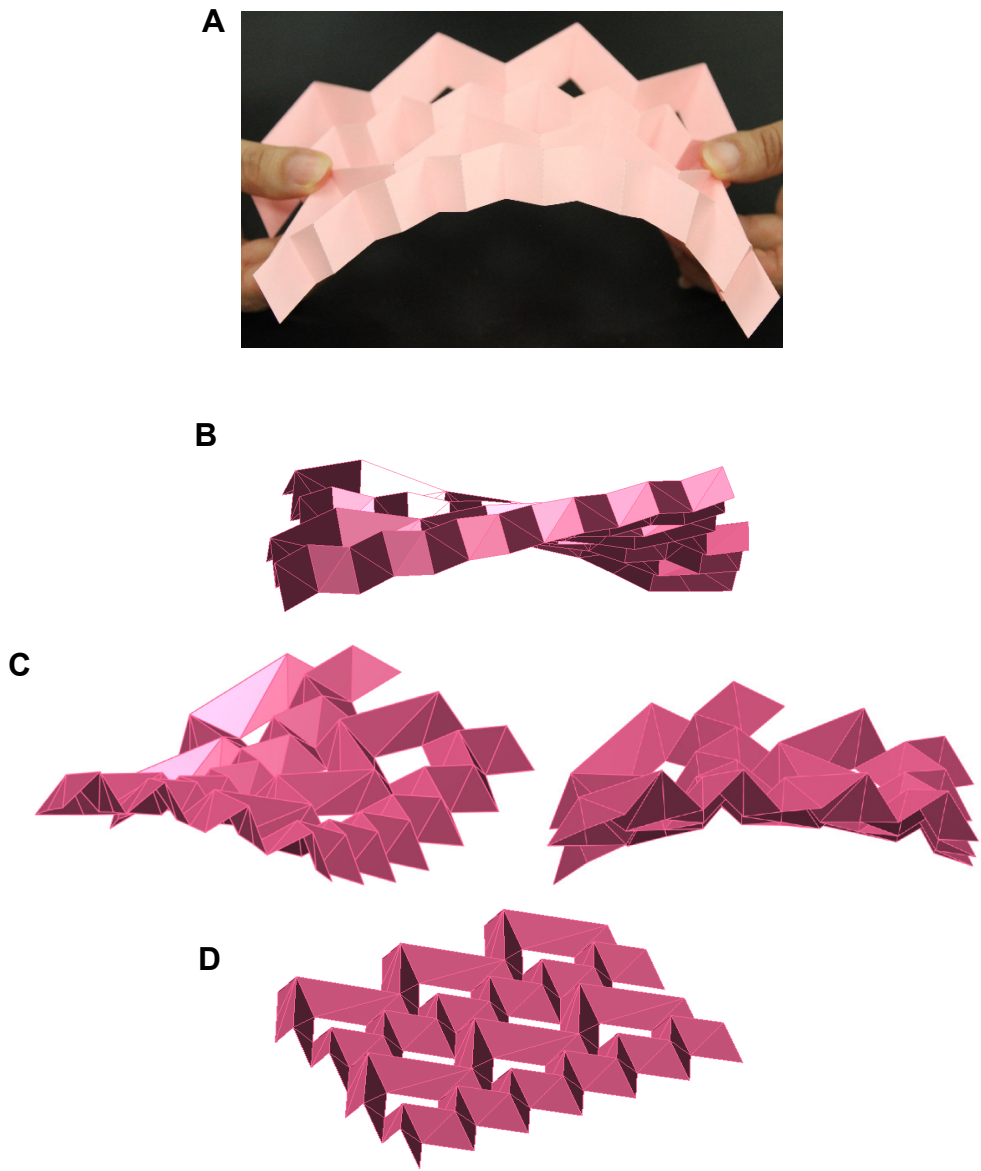

**Fig. A7. Behavior of a sheet of the pattern shown in Fig. 3C under bending and results of eigen-value analysis of a 2 by 3 sheet of the pattern.** (**A**) The sheet deforms into a saddle-shaped under bending (*i.e.*, typical behavior seen in materials having a positive Poisson's ratio). (**B**) Twisting, (**C**) saddle-shaped from two different views and (**D**) rigid origami behavior (planar mechanism) of a 2 by 3 pattern shown in Fig. 3C (with *a=1*; *b=2*; $\alpha = 60°$).

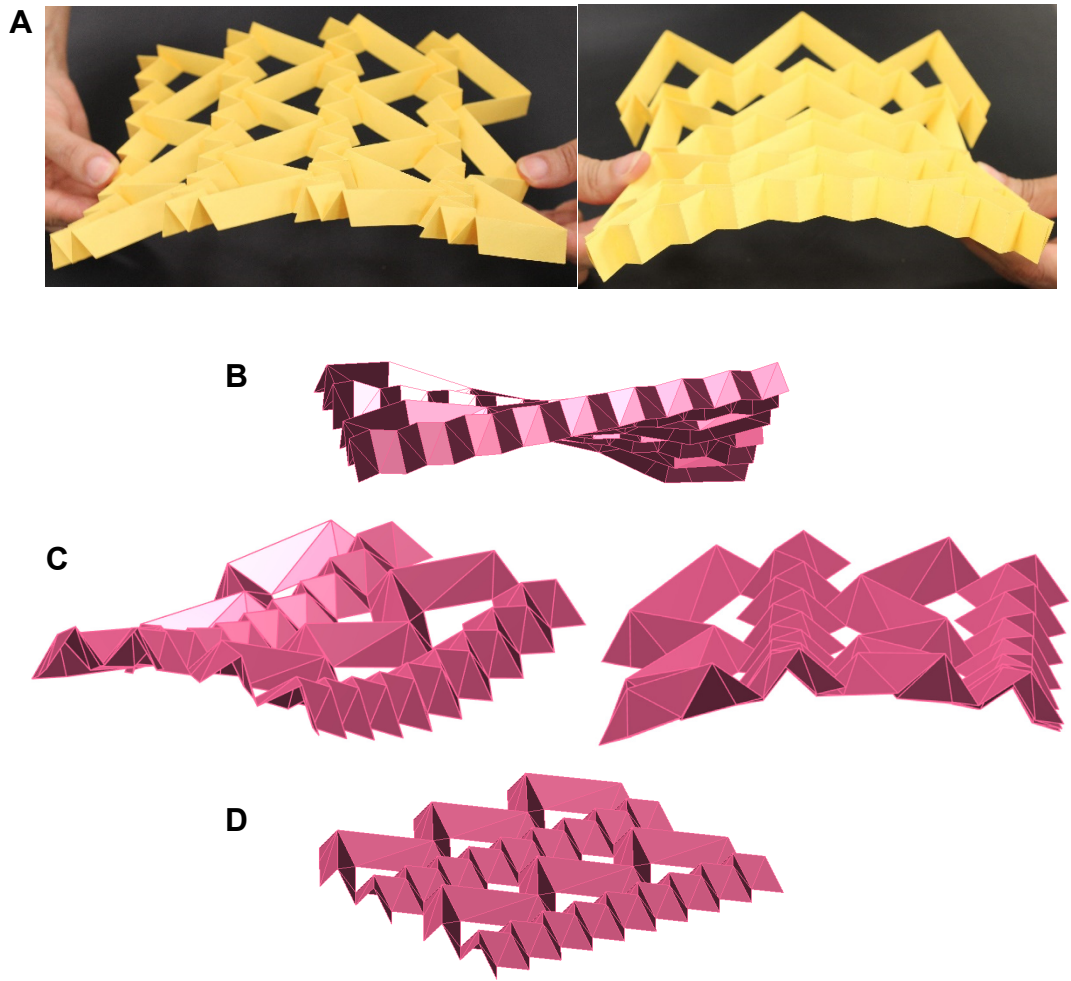

**Fig. A8. Behavior of a sheet of the pattern shown in Fig. 3D under bending and the results of eigen-value analysis of a 2 by 3 sheet of the pattern.** (**A**) The sheet deforms into a saddle-shaped under bending, *i.e.* a typical behavior seen in materials having a positive Poisson's ratio. (**B**) Twisting, (**C**) saddle-shaped from two different views and (**D**) rigid origami behavior (planar mechanism) of a 2 by 3 pattern shown in Fig.1 G (with *a=1*; *b=2*; $\alpha = 60°$ ).